\documentclass[12pt]{article}

\usepackage{graphicx}
\usepackage{amssymb}
\usepackage{amsthm}
\usepackage{booktabs}
\usepackage{rotating}
\usepackage{multirow}
\usepackage{eurosym}
\usepackage{amsfonts}
\usepackage{amsmath}
\usepackage{umoline}
\usepackage{caption}
\usepackage{subcaption}
\usepackage{color}
\usepackage{url}
\usepackage{authblk}
\usepackage[english]{babel}

\usepackage{pbox}
\usepackage{footnote}
\usepackage{multirow}
\usepackage{changepage}
\usepackage{tablefootnote}
\usepackage{ccicons}

\usepackage[margin = 2cm]{geometry}

\usepackage{hyperref}

\usepackage[natbibapa]{apacite}
 
\begin{document}

\title{Multilayer Network Analysis for Improved Credit Risk Prediction \footnote{\scriptsize NOTICE: This is a preprint of a published work. Changes resulting from the publishing process, such as editing, corrections, structural formatting, and other quality control mechanisms may not be reflected in this version of the document. Please cite this work as follows: \'{O}skarsd\'{o}ttir, M. and Bravo, C. (2021). Multilayer network analysis for improved credit risk prediction. \textit{Omega}, 105, 102520. DOI: \protect\url{https://doi.org/10.1016/j.omega.2021.102520}. This work is made available under a \href{https://creativecommons.org/licenses/by-nc-nd/2.0/}{Creative Commons BY-NC-ND license}. \ccbyncnd}}

\author[1]{Mar\'{i}a \'{O}skarsd\'{o}ttir}
\author[2]{Cristi\'{a}n Bravo}

\affil[1]{Department of Computer Science, Reykjav\'{i}k University, Menntavegur 1, 102 Reykjav\'{i}k, Iceland.}

\affil[2]{Department of Statistical and Actuarial Sciences, The University of Western Ontario,1151 Richmond Street, London, Ontario, N6A 3K7, Canada.}

\date{}

\maketitle

\begin{abstract}
    We present a multilayer network model for credit risk assessment. Our model accounts for multiple connections between borrowers (such as their geographic location and their economic activity) and allows for explicitly modelling the interaction between connected borrowers. We develop a multilayer personalized PageRank algorithm that allows quantifying the strength of the default exposure of any borrower in the network. We test our methodology in an agricultural lending framework, where it has been suspected for a long time default correlates between borrowers when they are subject to the same structural risks. Our results show there are significant predictive gains just by including centrality multilayer network information in the model, and these gains are increased by more complex information such as the multilayer PageRank variables. The results suggest default risk is highest when an individual is connected to many defaulters, but this risk is mitigated by the size of the neighbourhood of the individual, showing both default risk and financial stability propagate throughout the network.
\end{abstract}


\section{Introduction}

Network science has emerged as a fundamental theory and an essential tool to understand, describe, model and analyze the complex systems of interacting entities that arise naturally in e.g., biology, sociology, finance and economics, among many others \citep{barabasi2016network}.
In real life, objects are often connected by more than one type of relationship, for example in online social networks, two people can be friends on Facebook as well as follow each other on Twitter. 
Such networks are generally called multilayer networks, for which theory and model development has been a focal point within network science for the last decade.

In banking, it has long been suspected that correlated default exists. However, all studies that have researched propagation of risk across networks have focused on contagion across financial institutions \citep{bookstaber2016, feinstein2019}, or on the asset correlation with the overall economy first proposed in the Basel accords \citep{BCBSExplanatory2005}. Several authors have pointed out that these correlation effects do affect current credit risk measurements \citep{fenech2015, Thomas_2005} at an individual level, a fact that has been neglected so far. Given the importance of retail lending to the overall economy, and current advances in network science, we combine credit risk with multilayer network science to present a methodology to explicitly calculate the impact of correlated default in connected borrowers at a one-to-one level. Our proposal then targets not only understanding default propagation better, but also improving application scoring systems, commonly used to make loan granting decisions.

This paper has three main contributions. Firstly, we present a framework for creating a multilayer bipartite network from a tabular dataset with two or more connector variables. These variables are not necessarily explicit network variables, but can instead be a shared property or a characteristic.
Secondly, we develop a novel approach for credit risk prediction with a personalized multilayer PageRank centrality measure which ranks the nodes in the multilayer network with respect to a set of source nodes that can be used to understand correlated risk and also to improve credit scoring systems.
Finally, we show that influence scores from multilayer networks contribute to better performing credit scoring models.
While our results are general and applicable to any lending dataset (and to any problem in which correlated effects are suspected), we motivate the work in an environment in which correlated default is obvious. We focus our attention to multilayer networks in agricultural lending and present general algorithms to include them in any predictive model where they arise. As much as multilayer networks have been studied in banking, so far --and to the best of our knowledge-- its effects on retail lending (traditional consumer loans and over the counter loans to small and medium businesses) has not received the same level of attention. This is not surprising given its structure: retail loans are far more numerous, thus any network interactions require complex analysis that have only became feasible during the last decade. That said, it is also natural to expect borrowers to be connected in many complex ways, so the analyses used to measure their behaviour should also follow this level of complexity.
    
In this paper, we present a methodology for credit risk prediction, that continues the research direction started by \citet{oskarsdottir2019} and \citet{bravo2020a}. In these previous works, information from single layer social networks was shown to enhance the predictive performance of credit scoring models in retail lending, and there was an indication of correlation between the influence propagated through bipartite multilayer networks and default behaviour, respectively. 
In this current work, we build bipartite multilayer networks, which we extend by including the strength of interconnections between layers that we deem \emph{stickiness}.
We use a real agricultural lending dataset, an area of application that has been recognized as having strong network effects in their supply chains \cite{behzadi2018}, and supply chains have been show to improve when efficient credit is available \citep{zhong2018}. We connect borrowers, i.e. the producers, to the product which they produce and to the district in which they live to create multilayer networks.
We apply our novel personalized PageRank algorithm for multilayer networks with known defaulters as the source of influence and subsequently extract information from the network in terms of variables which we add to the lending dataset.
Then, to understand the predictive power of variables derived from such multilayer networks, we benchmark them against credit scoring models built with no network variables and with single layer models. We finally explore the new knowledge arising from these variables by using the SHapley Additive exPlanations \citep[SHAP;][]{shap2017} over the best performing Stochastic Gradient Boosting model, in order to interpret the new variables and understand what novel knowledge do they bring to the practice of credit risk assessment. 
Our results show a significantly enhanced model performance when including PageRank derived multilayer network variables, and a complex, non-linear, interaction between the variables. 

The structure of this paper is the following: the next section positions this paper in current literature, both in network science and in retail lending. The following section proposes the Multilayer Personalized Pagerank Algorithm which will be the basis of this work. Section \ref{sec:Experiments} presents the data set and the experimental results. The final section shows the conclusion of the paper.

\section{Previous works}\label{sec:Literature}
 
\subsection{Centrality in multilayer networks}
In complex systems, the type of interactions between entities in different subsystems can vary and thus multilayer network theory has evolved within network science in the last few years to model such interactions  \citep{kivela2014multilayer,barabasi2016network}.
In addition to the nodes and edges that are the fundamental components of regular single layer networks --or graphs-- multilayer networks have various layers to account for different types of interactions. 
In the most general case of multilayer networks, any node can belong to any subset of layers and an edge connects any two nodes in any layer or between any two layers.
Different forms of multilayer networks exist, depending on the characteristics of nodes and edges within and between the various layers and the constraints which they are subjected to. 
For example, in multiplex networks, inter layer edges only exist between the same node in all layers \citep{kivela2014multilayer}.

Centrality is an important central theme in network science, where some indicators of centrality are used to identify the most important nodes in a network.
The most common centrality measures are degree, closeness, betweenness and eigenvector --or PageRank-- centrality.
Degree counts the number of edges incident to a node and is both conceptually and computationally the simplest.
Closeness measures the average distance of a node to all other nodes in the network and thus how close it is to the other nodes.
Betweenness represents how often a node lies on the shortest path between the other nodes in the network.  
Closeness and betweenness are computationally expensive, since they require finding the shortest paths between all node pairs in the network.
PageRank centrality measures the influence a node has in a network. It depends on the number of nodes that link to the node and their PageRank centrality. It was originally designed to rank webpages in search engines \citep{page1999pagerank} but has been utilized in numerous applications \citep{kwak2010twitter,lohmann2010eigenvector,botelho2011combining, min2018behavior}. It is based on the stationary distribution of a random walker in a network who can randomly teleport to any other node by means of a restart vector. The restart vector of the equation can be manipulated to steer the random walk towards a set of source nodes. This is known as Personalized PageRank \citep{page1999pagerank}. It has been used in various applications to rank nodes with respect to source of influence so that the ones that are closer to the source obtain a higher score \citep{van2017gotcha}.

As the research on multilayer networks expands, so does the literature on the generalization of centrality measures, in particular PageRank-like centrality.
Because of the multiple types of edges and the complexity of additional layers, this is not a straight forward task.
One the one hand, several papers develop a PageRank centrality measure for multiplex networks.
\citet{halu2013multiplex}, for example, extended the PageRank centrality measure to multiplex network by computing a regular PageRank in one layer of the network and using the resulting ranking as input for the ranking on the second layer. 
Similarly, \citet{pedroche2016biplex} generalized the notation of the classical PageRank in the style of biplex networks and then extended them to multiplex networks.
Functional Multiplex PageRank is capable of capturing non-linear effects caused by different types of links between nodes \citep{iacovacci2016extracting, iacovacci2016functional} whereas
\cite{tu2019multiplex} defined a coupling Multiplex PageRank that can describe the coupling effects given to every distinct pattern of connections.
On the other hand, a multilayer PageRank centrality measure has been developed within MuxViz, a framework for the analysis and visualization of multilayer networks \citep{de2015muxviz}.
The framework includes extensions and mathematical formulation of several network concepts for multilayer networks such as  diffusion and centrality \citep{de2013mathematical, gomez2013diffusion, kivela2014multilayer, domenico2013centrality}. 
In the MuxViz framework, multilayer PageRank is computed similarly to the regular PageRank, but using eigentensors and aggregating the values of nodes that appear in more than one layer. 
\citet{cheriyan2020improved} extended this version of the multilayer PageRank measure, by introducing weights for the layers in the network. 

To summarize, much of the research to date has focused on generalizing the PageRank algorithm for multilayer networks. One of the contributions of this paper is a personalized PageRank for multilayer networks, which to the best of our knowledge, has not been introduced before.

\subsection{Learning from Networks}
The goal of this paper is to predict creditworthiness using network information from multilayer networks. 
Generally, this is a binary classification problem, where each observation --or node-- belongs to one of two classes, in this case default or not default.
Information from networks can be obtained in several ways. Firstly, with feature extraction, predefined variables that describe the nodes' neighbourhood and/or structural properties are computed for the nodes in the network, thus increasing the number of node attributes. The node attributes can then be used together with a supervised learning technique to predict creditworthiness. 
Secondly, representation learning is a technique to automatically obtains a feature embedding for the nodes in the network. The embedding can subsequently be used as input for binary classifiers \citep{grover2016node2vec}.
Finally, classes of nodes can be learned directly from the network, using network learning techniques \citep{macskassy2007classification} or deep learning techniques such as graph neural networks.

All approaches have been used successfully for regular single layer networks in various applications.
Feature extraction was used to predict churn in telco \citep{oskarsdottir2017}, for social security fraud detection \citep{van2017gotcha} and credit scoring \citep{oskarsdottir2019}.
Representation learning has been used for a variety of tasks, including link prediction \citep{grover2016node2vec}, and node labelling such as churn prediction \citep{oskarsdottir2019inductive}.
Network learning with deep learning approaches is currently on the rise \citep{bui2018neural, zhou2018graph}.

However, the literature does not extend to multilayer networks apart from representation learning.
Representation learning recently emerged as a technique to map vertices and information about their features and structural properties into a lower dimensional vector space \citep{zhang2018network}. 
As a result, the network is easy to handle for subsequent analyses, such as link prediction, node classification and community detection.
At first research efforts were mainly dedicated to single layer networks, but in the last couple of years the focus has shifted to the development of representation learning for multiplex networks \citep{bagavathi2018multi,li2018multi,ning2018representation,zhang2018scalable,cen2019representation}. Although such techniques are capable of preserving both structure and content of nodes, as well as being scalable to large networks, their main downside, especially in the context of lending, is the lack of interpretability of the learned embeddings. 

Some research on feature extraction of multilayer networks exist. For example, \citet{amoroso2018multiplex} used multiplex networks based on MRI images to detect early signs of Alzheimer's. They featurized the networks and used random forests to predict signs of the disease.In our approach, we similarly, extract features from multilayer networks and study their predictive performance in terms of credit risk.

We can see that most of the existing research on learning from networks utilize single layer and unipartite networks. 
We extract features which we design based on multilayer networks and the outputs of the propagation algorithm. 

\subsection{Network Effects in Retail Credit Risk Measurement}

As we mentioned before, network connections have been shown to affect credit risk. Connections between banks have been linked to the systemic risk of an economy \citep{thurner2013debtrank}. This study however focused on the connection between the companies that participate in a financial system, not on the borrower themselves, which are the customer of these entities. A similar, but now multilayered, analysis was conducted by \citet{montagna2016}, who identified that default contagion risks across financial firms are non-linear. And finally, very recently \citet{gupta2020constrained} studied the use of clustering methods to identify bipartite network effects in banks in India. All of these works strongly suggest that network effects can increase the risk of the system as a whole.

On the borrower side, our previous work \citep{oskarsdottir2019} showed how connectivity between borrowers across a single network affects credit risk. Network variables were some of the stronger predictors of default over a credit card data set. Our previous work, however, did not dwell on multilayer effects, which have been shown relevant to estimate risk. In \citet{poledna2015multi}, the authors have pointed out that systemic risk estimates can be severely underestimated if multilayer effects are ignored. This latter study focused again on a financial institutional network instead of a borrower-level network.

A more recent study by \citet{cheng2020} focuses directly on networked loans. These loans are granted not to one individual, but to a clique in which all users are guarantors of each other. Their results suggest, not surprisingly, that the composition of the clique is extremely important on the default risk of each individual. The effects in this work are based on the explicit network of guarantees though, without going further into other types of connections.
Being aware of the complexities of credit risk evaluation, \citet{bai2019banking} develop a method with fuzzy rough sets in agricultural lending, in order to extract and investigate the complex relationships between farmer characteristics,  competitive environmental factors, and the farmers' creditworthiness. 

In conclusion, previous literature on network effects on credit risk are  focused on either the financial institutions that compose the market, as opposed to the borrower level, or on a single network instead of a multilayer one. In this work we tackle the more complex, and more general, problem of measuring risk across multilayer networks at the borrower level, and then constructing a predictive model out of the outputs of the network analysis.

\section{Personalised Multilayer PageRank Centrality}\label{sec:Methodology}

\subsection{Bipartite Multilayer Network Construction}
\begin{figure}
    \centering
    \includegraphics[width=1.0\linewidth]{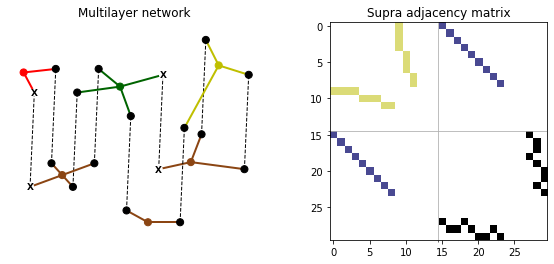}
    \caption{A multilayer network (left) and its supra adjacency matrix (right). The multilayer network has nine borrower nodes (black), three location nodes (brown) and three product nodes (red, green, yellow) in two layers. The nodes in the first layer are connected with brown edges and the nodes in the second layer are connected with red, green and yellow edges. The inter edges (darkgray) connect a borrower node to itself in the other layer. In the supra adjacency matrix the upper left submatrix (yellow) denotes the adjacency matrix for the first layer, the lower right matrix (black) is the adjacency matrix for the second layer. The upper right and lower left submatrices (blue) are the adjacency matrices for the inter edges. They are diagonal as each node is only linked to itself in the other layer. The black nodes marked with \textbf{X} are the source of influence.}
    \label{fig:ExMul}
\end{figure}

Let $G=(V,E)$ be a single layer network --a graph-- where $V$ is a set of nodes and $E\subseteq V\times V$ is a set of edges that connect pairs of nodes. 
A network is represented by its adjacency matrix or rank-2 adjacency tensor $W_j^i$ which encodes information about the connections from node $i$ to node $j$ and their intensity in terms of edge weights. If the network has $N=|V|$ nodes then $W_j^i$ has dimension $N\times N$.

In a \emph{bipartite network} $G=(V_1,V_2,E)$, each node belongs to one of two independent and disjoint sets, $V_1$ or $V_2$, and edges exist only between nodes in different sets, that is $E\subseteq V_1\times V_2$.
Let $N_1=|V_1|$ and $N_2=|V_2|$. The adjacency matrix of a bipartite network is given by the $(N_1+ N_2)\times (N_1+ N_2)$ matrix where entries in the first $N_1\times N_1$ and the last $N_2\times N_2$ blocks are zero since no edges exist between nodes in the same set.

Networks with more than one type of edges are represented using multiple layers where one type of edges connects the nodes in each layer. These networks are called \emph{multilayer networks}.
The nodes in each layer are connected based on one type of relationship, so the various layers are used to account for all different types of relationships. Edges can exist between any pair of nodes in the same layer, or between any pair of nodes in any pair of layers.  These are called intra and inter edges, respectively.

A multilayer network $\mathcal{M}$ with $N$ nodes and $L$ layers can be represented with an $N\times N\times L\times L$ dimensional rank-4 adjacency tensor $M_{j\beta}^{i\alpha}$ which indicates the weight of an edge between node $i$ in layer $\alpha$ and node $j$ in layer $\beta$ \citep{de2013mathematical}. By convention, we represent nodes with Latin letters and layers with Greek letters.
To simplify notations and computations, the rank-4 adjacency tensor can be flattened to achieve an equivalent representation of the network as a rank-2 tensor which is an $N \cdot L \times N \cdot L$ matrix, known as supra adjacency tensor or matrix.

In this paper we consider multilayer networks with a bipartite network in each layer. 
We assume that there is a set of nodes that exist in all layers and call them \emph{common nodes}, denoted with $V_c$ with the number of common nodes as $|V_c|$.  
We furthermore assume, that each layer has its own set of \emph{specific nodes} that only exist in that specific layer.
Given $L$ layers $\alpha, \beta, \dots \omega$, let the set of specific nodes in each layer be $V_s^{\alpha}, V_s^{\beta} \dots V_s^{\omega}$, with the number of specific nodes in each layer given by $|V_s^{\alpha}|,| V_s^{\beta}|, \dots, |V_s^{\omega}|$.
When constructing the supra adjacency matrix of these types of networks, we let $N=|V_c|+|V_s^{\alpha}|+| V_s^{\beta}|+\dots +|V_s^{\omega}|$ be the number of distinct nodes in the network. This means that all specific nodes exist in each layer, but only edges between the common nodes and the layer's specific nodes are allowed. This is to ensure that the dimensions of the matrices match.
In these multilayer networks, inter layer edges exist only between the common nodes, i.e. the black nodes in Figure \ref{fig:ExMul} where inter layer edges are represented by dashed lines. The weight on the inter layer edges gives an indication of how easy it is to travel between layers, with a higher weight indicating a more likely transition.  We refer to this inter layer weight parameter as \emph{stickiness}, $S$. The default value for $S$ is 1, meaning it is unweighted. Note that as $S\to0$, the connection between layers gets weaker and weaker until the layers end up being disconnected. In contrast, as $S\to\infty$, the layers get aggregated, resulting in a network with a single layer but several types of edges, i.e. an edge-coloured network. 

The subfigure on the left of Figure~\ref{fig:ExMul} shows an example of such a multilayer network with two layers. There are nine common nodes (black), three nodes which only belong to the bottom layer (brown) and three nodes which only belong to the top (red, green or yellow).

For ease of notation, we assume without loss of generality, that there are two layers, $\alpha$ and $\beta$, with intra layer adjacency matrices $A_j^i$ and $B_j^i$, the common nodes $V_c$, and the specific nodes $V_s^{\alpha}$ and $V_s^{\beta}$ in layers $\alpha$ and $\beta$ respectively. Then the supra adjacency matrix $M_{j\beta}^{i\alpha}$ can be written as
\[
M_{j\beta}^{i\alpha}=\left [\begin{array}{c|c}
    A_j^i & I_i^i \\ \hline
    I_i^i & B_j^i
\end{array} \right]
\]
where the matrices $A,B,I$ have dimension $N\times N$ with $N=|V_c|+|V_s^{\alpha}|+|V_s^{\beta}|$ and
 $I$ is zero except for values on the diagonal corresponding to the common nodes to indicate $S$, the weight of the inter layer edges. The right subfigure in Figure~\ref{fig:ExMul} shows the flattened adjacency tensor for the multilayer network on the left.
 This can easily be extended for more than two layers, adding intra layer adjacency matrices to the diagonal of the supra adjacency matrix and inter layer adjacency matrices on the off diagonal, maintaining the symmetry.

To describe the transition probabilities of a random walker traversing the multilayer network, both within and between layers we compute the supra transition matrix $T_{j\beta}^{i\alpha}$, i.e. the column-normalized supra adjacency matrix \citep{garas2016interconnected}.

\subsection{Multilayer PageRank Centrality}\label{sec:multiplexpagerankc}
The PageRank centrality of a node is equivalent to the probability that a random walker who is traversing the network would end up at the node. 
In addition to travelling along the network's edges, the random walker can also jump --or teleport-- to some other node at random in the network.
In a single-layer network, this is the steady-state solution of the equation $p_j(t+1)=R_j^ip_i(t)$ where $R_i^j$ is the transition matrix of a random walker that travels to a neighbouring node with probability $r$ but jumps to any other node in the network with probability $1-r$. The parameter $r$ is called restart or damping factor. It determines the trade-off between the importance of the network and the random jump in the random walk.
Traditionally it is set to 0.85, since that value ensures fast convergence while accurately representing web browsing behaviour: a web surfer clicks on hyperlinks roughly five-sixths of the time but visits a new page one-sixth of the time \citep{langville2004deeper,boldi2007deeper,bressan2010choose}.

In a multilayer network, the random walk can traverse between nodes in the same layer as well as between layers along inter layer edges. Similarly, the walker can jump to any node in any layer.
Therefore, the transition tensor of the random walk becomes
\begin{equation}\label{eq:R}
    R^{i\alpha}_{j\beta}=rT^{i\alpha}_{j\beta}+\frac{1-r}{N \cdot L}u^{i\alpha}_{j\beta},
\end{equation}
where $T^{i\alpha}_{j\beta}$ is the rank-2 supra transition tensor of the network, $r$ is the damping factor and $u^{i\alpha}_{j\beta}$ is an $N \cdot L \times N \cdot L$ matrix of ones, to indicate equal probability of jumping to any node in the multilayer network.

The transition tensor $R^{i\alpha}_{j\beta}$ represents the random walk, i.e. the probability of jumping between pairs of nodes. Let $p_{i\alpha}(t)$ be a time dependent tensor that represents the probability of finding the walker in a given node and layer at time $t$. Then the random walk is 
\[
p_{j\beta}(t+1)=R^{i\alpha}_{j\beta}p_{i\alpha}(t).
\]
The multilayer PageRank centrality is the steady state solution to this equation, that is, when $t \to \infty$, and is given by $\Pi_{i\alpha}$, which represents the probability of finding the walker at node $i$ in layer $\alpha$. The solution is obtained by finding the leading eigentensor, which is equivalent to the solution of the higher-order eigenvalue problem,
\[
T_{j\beta}^{i\alpha}\Pi_{i\alpha}=\lambda\Pi_{j\beta}
\]
as derived in \citet{domenico2013centrality}.

The steady state solution $\Omega_{i\alpha}$ gives the probability of finding the walker at node $i$ in layer $\alpha$. To finally reach a value for the PageRank centrality for each node, the values in all layers are aggregated, which gives $\omega_i$: the multilayer PageRank centrality of node $i$.

\subsection{Influence Matrix}
The multilayer PageRank centrality derived above gives a representation of node importance assuming that the random walk jumps to any other node in any layer with equal probability.
However, in some situations it is beneficial to bring out nodes that are important from the perspective of a set of specific nodes. 
This is achieved by allowing the walk to jump to a set of specific nodes only, thus making the random walk biased. This version is called personalized PageRank \citep{page1999pagerank} and has for example been used to detect social security fraud \citep{van2017gotcha}.

Although PageRank has been derived for multilayer networks, to the best of our knowledge, multilayer personalized PageRank was derived for the first time by \cite{bravo2020a}.
Below, we describe how the multilayer PageRank can be generalized for biased random walks to obtain a personalized score.

\begin{figure}
    \centering
    \includegraphics[width=1.0\linewidth]{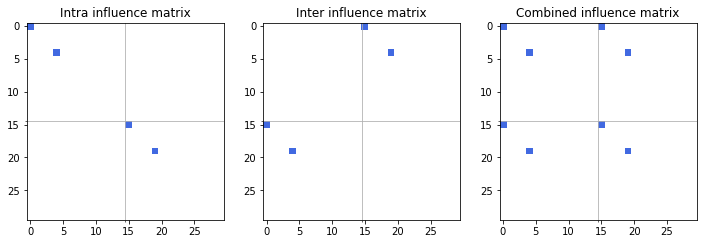}
    \caption{Three scenarios for the influence matrix of a two layer multiplex network when computing a personalized PageRank. }
    \label{fig:personalized}
\end{figure}

Starting with Eq. \ref{eq:R}, we modify the $u^{i\alpha}_{j\beta}$ matrix, by setting only specific values equal to one and the remaining values to zero, thus allowing the random walk to jump only to nodes in a specific set, which we call influence nodes or the source of influence, denoted with $V_I\subset V$.  We refer to the $u^{i\alpha}_{j\beta}$ matrix as \emph{influence matrix}.
However, setting all values in the influence matrix to zero except for those where influence originates is non-trivial, as these nodes appear several times in $u^{i\alpha}_{j\beta}$, both within and between layers. 
To illustrate this,  assume there are two layers, so that $u$ has dimension $2N\times 2N$ and can be split into four blocks of $N\times N$ sub-matrices. Two of those blocks are on the diagonal, $u_{11}$ and $u_{22}$, and correspond to the intra-layer edges; and the other two blocks are off the diagonal, in the upper and lower triangle, and correspond to the inter-layer edges. We call them $u_{12}$ and $u_{21}$. Then the matrix looks as follows
\[
u^{i\alpha}_{j\beta}=\left [\begin{array}{c|c}
    u_{11} & u_{12} \\ \hline
    u_{21} & u_{22}
\end{array} \right]
\].
We propose three scenarios to define the influence matrix.
\begin{enumerate}
\item
Intra-influence: In this case, the random walk jumps to $V_I$ nodes within the layers. We assign the value 1 to the $V_I$ nodes on the diagonal in the intra-layer matrices $u_{11}$ and $u_{22}$ 
\item
Inter-influence: In this scenario the influence originates between the layers. We assign the value 1 to the $V_I$ nodes on the diagonal of the two inter-layer matrices, $u_{12}$ and $u_{21}$
\item
Combined-influence: Here, influence originates both within and between the layers, thus we assign the value 1 to the $V_I$ nodes on the diagonal of all four submatrices in $u^{i\alpha}_{j\beta}$.
\end{enumerate}
Figure \ref{fig:personalized} shows these three scenarios for the influence matrix of the two layer network in Figure \ref{fig:ExMul} where the influence nodes are marked with a black X.

Finally, the denominator of the second term in Eq.\ref{eq:R} reduces to the sum of the elements in $u^{i\alpha}_{j\beta}$.
With a redefined transition tensor for the random walk, we proceed to compute the PageRank centrality as in section \ref{sec:multiplexpagerankc}.

\section{Experimental Results}\label{sec:Experiments}

\subsection{Dataset}
For this work, we use the data set previously presented in \citet{Bravo_2013}. The descriptive statistics of the data and the variable descriptions can be seen in~\ref{app:descriptive}. This data set focuses on agricultural lending in a Latin American country, with a set of over 70,000 loans of between one to five years in duration, covering 15 years of data (1998 - 2013). The data set includes information regarding the loan, such as the amount, the term, whether there are guarantors and/or collateral, and the value of the latter if present. It also includes sociodemographic information of the borrower, including the borrower's age, district in which they operate, the product(s) they trade in, and, if available, any information concerning past loans (number, amounts, terms, past arrears and other previous financial behaviour). The data has been previously cleaned of outliers, removing only extreme values which would be out of policy for the financial institution, and null values.

\subsection{Network setup, score calculation and variable extraction}
The goal of our case study is to complement the loan specific information in the data set with network information and specifically to quantify influence from loans that defaulted.
We achieve this by building multilayer networks, to which we apply the personalized PageRank algorithm to obtain influence scores as well as other network features. These features are then added to the loan data set, before building credit scoring models for the borrowers. Below we describe the process in detail.

First, we use the data to build a sequence of multilayer networks as follows.
Each multilayer network is created using the loans granted within a five year period \footnote{This period rendered networks with almost all nodes ($>99.5\%$) in the same connected component, which is essential to get stable results from the PageRank algorithm in the next step.}.
Each network consist of two layers. Which network layers should be used is a very important step in multilayer modelling. The chosen networks should be those that the modeller believes are the ones that propagate the studied effect (default in this case) for the individuals under study. In many circumstances, the available networks will be simple ``the ones that are available'', but in other circumstances, there may be many choices. Nothing precludes the potential modeller to include a large number of networks, but they must always keep in mind that the complexity of calculation grows with the square of the network size, as shown below.

For our experiments, we will use the geographical district and the product as the two layers of our model. Both have been recognized as some of the most important factors in agricultural lending, particularly in the small farmer segment \citep{peckchristen2012}. The first layer we use then is the product layer, where each borrower is connected to their respective product, thus multiple borrowers share a connection if they plant the same product. Note that farmers may plant more than one product per year (rotating crops), one farmer may be connected to more than one other farmer by using this product network.
The second layer is the district layer, where each borrower is connected to their respective district and to their respective area (several areas make up one district). In effect, we have set the stickiness parameter between the area and the district networks to $\infty$, thus aggregating the two layers. This was done since areas are subsets of districts, and hence very closely connected. As a result, this means that borrowers that live in the same area are more tightly connected than those that live in the same district only.
Each of the two layers thus consists of a bipartite network.
To complete the network construction, we connect each borrower with themselves in the two layers.
These inter-layer edges can be given weights to represent the cost of travelling between the two layers. This stickiness is one of the parameters we tune in Section \ref{sec:ParamTuning}.

Next, we apply the multilayer personalized PageRank algorithm to the network in order to compute influence scores.
The goal is to assess the influence that past borrowers have on new ones. We achieve this by including in $V_I$ all borrowers that defaulted during the first four years and eleven months of the five year period. They are the source of influence in the influence matrix.  We then apply the algorithm and inspect the scores which borrowers in the final one month of the five year period get. 
These borrowers and their scores are later used when building the credit scoring models. 

The Personalized PageRank scores depend on the restart parameter $r$. It controls the trade off between the effect of the network and influence from the source nodes when computing the score.  
$r$ can assume any value between 0 and 1, and therefore the tuning of this parameter is included in our experiments.

In addition to the personalized PageRank scores, we extract other information for the borrowers appearing in the final month of the five year multilayer networks, see Table \ref{T:features} in \ref{app:features}.
Inspired by \citet{getoor2005link} we count the number of borrowers and the number of defaulted borrowers, in the previous one and five years, in the nearest neighborhood of each node in both layers of the network and in their intersection. Thus we obtain the degree variables in Table \ref{T:features}. In addition we extract the score of the product/district/area to which a borrower is connected, that is, the specific node in each layer. If a borrower is connected to more than one specific node in the same layer we only consider the largest value. 

For each value of  $r$ and $S$, we run the multilayer personalized PageRank algorithm three times: once with the intra influence matrix, once with the inter influence matrix and once with the combined influence matrix.
As a result of this process, the borrowers acquire a set of network variables, which are listed in Table \ref{T:features} in \ref{app:features}.

In addition, we aggregate the layers and consider the case where $S\to\infty$, that is, where there is only one layer, but two types of edges: product and district. We compute the regular Personalized PageRank with known defaulters as the source of influence for various values of $r$.  This approach represents the naive way of applying personalized PageRank to the network of borrowers.

To obtain the network variables for all the borrowers in the data set, we repeat this process, in each step building a network spanning five years and computing the scores of the 'new' borrowers, that is, the ones in the final month, and then shifting the five year period by one month, thus looking at a new set of borrowers.
This allows us to assess the influence of prior defaulters on the new borrowers by adding their network variables to the data set.
Note that borrowers that got their loan in the first 4 years and eleven months of our observation period must be excluded from the credit scoring exercises, since we do not have sufficient information about prior defaulters for them.

\subsection{Predictive Models Setup and Parameter Tuning}\label{sec:ParamTuning}

\begin{figure}[t]
\centering
\begin{subfigure}{0.5\textwidth}
\centering
    \includegraphics[width=\linewidth]{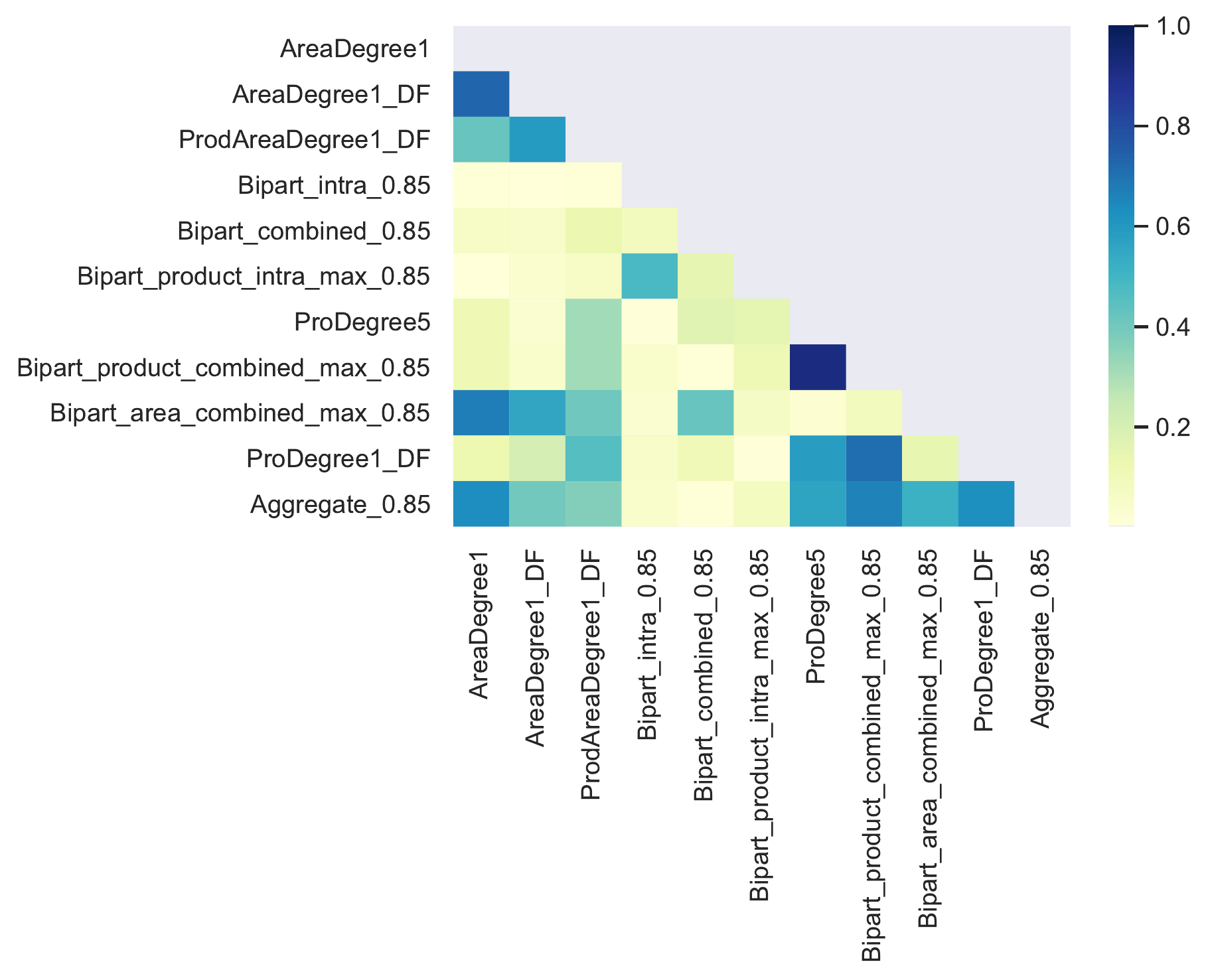}
    \caption{Network variables}
    \label{fig:CorrNetwork}
\end{subfigure}%
\begin{subfigure}{0.5\textwidth}
\centering
    \includegraphics[width=\linewidth]{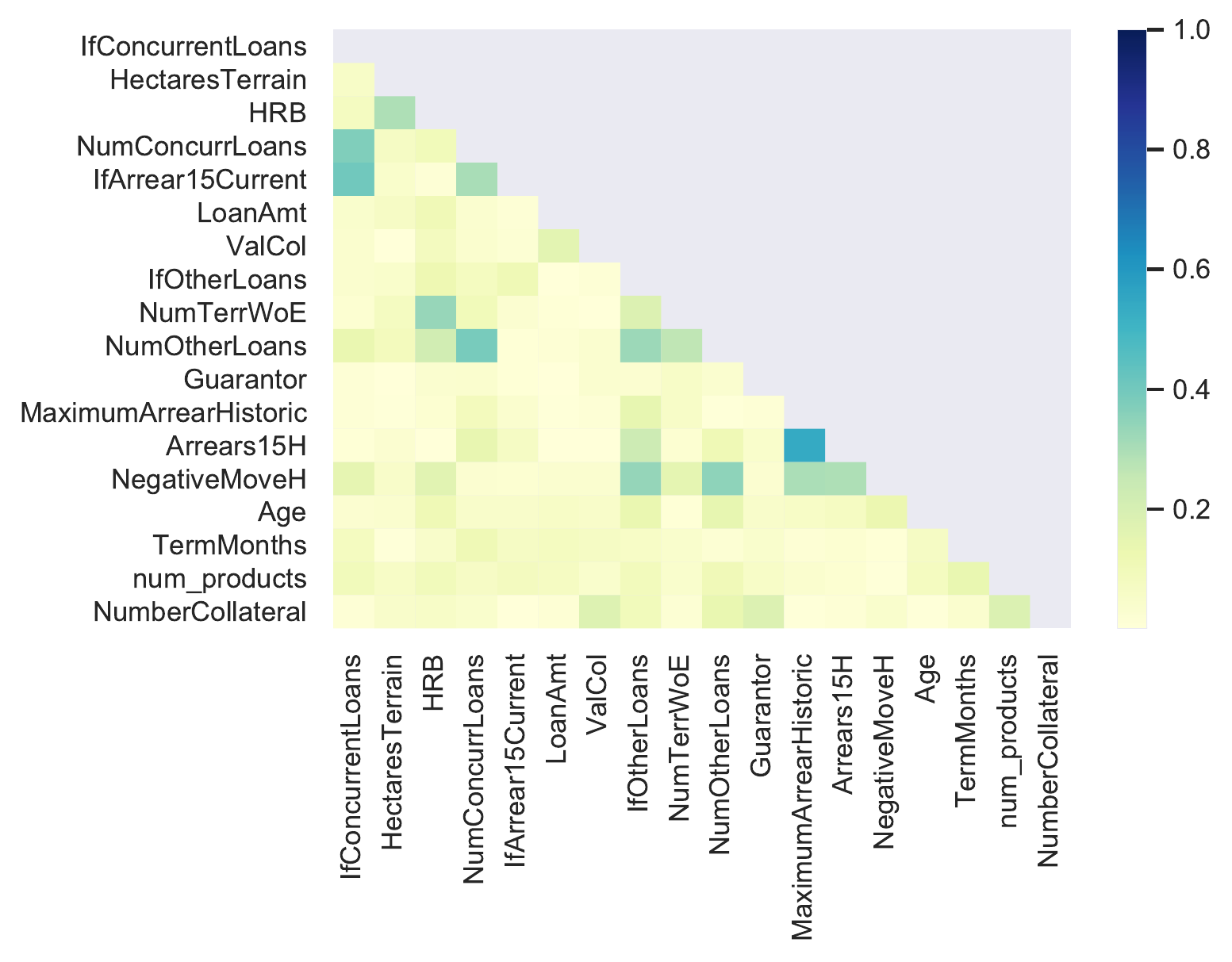}
    \caption{Non-network variables}
    \label{fig:CorrNonNet}
\end{subfigure}
\caption{Correlation matrices for both final sets of variables.}
\label{fig:Corrs}
\end{figure}

Before running the model we first remove highly correlated variables, setting the cutoff at 70\% correlation given there was a U-shaped distribution of the correlations, with a group below 30\% correlation, almost none in the 30\%-70\% range, and the rest above the cutoff of 70\%. In general, if making a choice between correlated variables, we chose the most correlated one with the target. From the original set of 32 network variables (see Table \ref{T:features} in \ref{app:features}), only seven are uncorrelated with other variables. In particular, most ProdDegree variables are correlated with the combined ProdAreaDegree variables, except for ProdDegree5, the long term centrality associated with the product network. All District variables are, as expected, correlated with the Area variables, but the latter are more informative (measured by their correlation with the default variable) so we keep only them in the model. Another interesting result is that there is a high correlation between almost all of the one year network variables and their five year counterparts, but the one year variables are better correlated with Default (except for the centrality for the product network, ProdDegree5). This means that, for this particular data set, short term effects are more important than long term ones. This will not necessarily replicate in other data sets, so we recommend users to perform the same analysis we have just performed. Finally, when selecting which versions of the influence matrix to use, we saw they tended to be very correlated, and therefore we decided to use the ``combined'' variable, since it contains the information of the combination of the inter and intra versions, plus the combined multilayer effect. This means one of the latter variables are redundant. The surviving variable, however, is highly correlated with the pagerank score of the aggregated network (Aggregate). This means only the multilayer scores and the flattened scores remain in the data set, as they are the most informative of the set. Figure~\ref{fig:Corrs} shows the final correlated matrices, differentiating between the network and non-network variables. Note there is no correlation between variables of the two groups.

We follow the process by benchmarking two contrasting models. Logistic regression is the industry standard, accounting for more than 95\% of all models used in banks \citep{thomas2017}, due to its simplicity and transparency. On the other hand, XGBoosting has been shown to be a powerful tree ensemble, reaching state-of-the-art performance in most credit risk structured data sets \citep{lessmann2015}.

For the logistic regression model we use LASSO penalization to reduce the number of meaningless variables, which is particularly important given we are benchmarking network variables. We tune the $\lambda$ parameter (the penalization weight) by using a three-fold cross-validation procedure over a 20\% subset of the training set. The XGBoosting on the other hand has a larger set of parameters that need to be tuned. We tune, for each subset of variables, the number of trees ($n\_estimators \in [50, 100, 250, 500]$ using the well-known Python/R package \emph{xgboost} notation), the maximum depth of each tree ($max\_depth \in [2, 3, 4]$), the learning rate ($learning\_rate \in [0.001, 0.01, 0.1]$), and the regression regularizer ($reg\_alpha \in [0.1, 0.2, \ldots, 1]$). We again use a subset of 20\% of the training data and a three-fold cross-validation procedure. Each model, built over different subsets of variables, then uses a (distinct) set of optimal parameters with the purpose of isolating the impact of the variables themselves instead of simple parameter choice.

The final set of parameter optimization that was performed considered the networks themselves. As discussed in section~\ref{sec:multiplexpagerankc}, there are two parameters that affect the network variables: the parameter $r$ which controls the trade-off between the adjacency matrix and the influence matrix $u^{i\alpha}_{j\beta}$, and weight of inter-layer edges which determines the cost of travelling between layers, the \emph{stickiness} $S$. Both parameters were tuned by selecting the best AUC (Area Under the Receiver Operating Characteristic Curve) for the model with full variables. As the model with all variables has the largest amount of information, and both the LASSO-regularized logistic regression and the XGBoosting model come with variable selection (regularization) capabilities, we believe this leads to the best parameters for all other experiments, as they run in subsets of the full dataset. The stickiness had little influence over the results, with significant worsening of the AUC values across the board for $S=0.25$ and $S=4$, but no effect elsewhere in the data set. We thus recommend using neutral stickiness (i.e. $S=1$) in the network construction, which assumes an entity is equally likely to stay in their own network than crossing to another layer. The $r$ value has a more interesting behaviour. Table~\ref{tab:alphas} shows the AUC value over the validation set for different values of $r$. The conclusion is that a high value of $r$ (i.e. a high weight on the adjacency matrix) is necessary to optimize the performance of the network variables. In our experiments the values of 0.8 and 0.85 perform equivalently, so we choose the value of 0.85, proposed in the original PageRank paper.

\begin{table}[tbp]
\centering
\caption{AUC values for different values of the $r$ parameter over the validation set,  all variables.}
\label{tab:alphas}
\begin{tabular}{cc}
\toprule
$r$ & AUC (validation set) \\ \midrule
0.2   & 0.733                \\
0.3   & 0.737                \\
0.4   & 0.734                \\
0.5   & 0.734                \\
0.6   & 0.734                \\
0.7   & 0.738                \\
0.8   & 0.738                \\
0.85  & 0.738                \\
0.95  & 0.735               \\
\bottomrule
\end{tabular}
\end{table}

With all parameters set, we can now discuss the model performance across different sets of variables.

\subsection{Prediction results}

\begin{table}[tb!]
\centering
\caption{AUC for the different models, and difference against the respective baseline model (with or without behavioural variables). The best performing model in each group is in bold.}
\label{tab:AUC_Summary}
\scriptsize
\begin{tabular}{lcc|cc}
\toprule
                                                     & \multicolumn{2}{c}{Logistic Regression}                                      & \multicolumn{2}{c}{XGBoosting}                                               \\
$r$ = 0.85, $S$ = 1                         & \multicolumn{1}{c}{AUC Test} & \multicolumn{1}{c}{\% Increase} & \multicolumn{1}{c}{AUC Test} & \multicolumn{1}{c}{\% Increase} \\
\midrule
Baseline                                             & 0.639                        &           -                                   & 0.660                        &  -                                             \\
Baseline + Centrality   (Product)                    & 0.656                        & 2.7\%                                         & 0.707                        & 7.1\%                                         \\
Baseline + Centrality (Area)                         & 0.695                        & 8.8\%                                         & 0.719                        & 8.9\%                                         \\
Baseline + Centrality (single   networks)            & 0.698                        & 9.2\%                                         & 0.733                        & 11.1\%                                        \\
Baseline + Centrality (single   + multilayer)         & 0.698                        & 9.2\%                                         & 0.729                        & 10.5\%                                        \\
Baseline + Personalized   PageRank                   & 0.648                        & 1.4\%                                         & 0.695                        & 5.3\%                                         \\
Baseline + All Network   Variables                    & \textbf{0.703}     & \textbf{10.0\%} & \textbf{0.737} & \textbf{11.7\%}                                        \\
\midrule
Baseline + Behaviour                                 & 0.788                        &         -                                     & 0.799                        &            -                                   \\
Baseline + Behaviour +   Centrality (Product)        & 0.794                        & 0.8\%                                         & 0.818                        & 2.4\%                                         \\
Baseline + Behaviour +   Centrality (Area)           & 0.802                        & 1.8\%                                         & 0.818                        & 2.4\%                                         \\
Baseline + Behaviour +   Centrality (single network) & 0.805                        & 2.2\%                                         & \textbf{0.825}                        & \textbf{3.3\%}                                         \\
Baseline + Behaviour +   Centrality (all)            & 0.805                        & 2.2\%                                         & 0.824                        & 3.1\%                                         \\
Baseline + Behaviour +   Personalized PageRank       & 0.791                        & 0.4\%                                         & 0.814                        & 1.9\%                                         \\
All Variables including   Behaviour                  & \textbf{0.807} & \textbf{2.4\%} & \textbf{0.826}                        & \textbf{3.4\%}                                        \\
\bottomrule
\end{tabular}
\end{table}

Table~\ref{tab:AUC_Summary} shows the performance of the models measured in terms of AUC. The models were built with different combinations of variables. As behavioural variables, if available, tend to be  most important in credit risk \citep{Bravo_2013} we test independently sociodemographic and loan data sets (Baseline variable set) from the Behaviour variable set. The network variables are divided given their complexity and layer of origin. The Centrality set has four subsets: Area and Product (degree and default degree centrality calculated only over either the area and product layer), single networks (both variables added to the model, calculated over the single layers), and single + multiplex (both the single centrality plus an additional centrality variable calculated over the multilayer network). 
The centrality measures are the degree and the default degree, i.e. the number of defaulters in the neighbourhood, during the previous one or the previous five years. 
The second group are the personalized PageRank variables, calculated over the multilayer network using the process outlined in Section~\ref{sec:multiplexpagerankc}. The final subset includes all variables in the network group. In total, we tested 14 different variable subsets, across both benchmark models, totalling 28 experiments. All reported AUC values come from an independent holdout test set.

The results show that the models including all network variables are significantly better than the both baseline models (with and without behaviour). This difference is reduced as expected for the models including behavioural variables (3.4\% improvement in the All Variables XGBoosting model vs 11.7\% for the Baseline + All Network model), and this gain is consistently higher for the XGBoosting model than for the Logistic Regression model, across all variable subsets. This suggest network variables have strong, non-linear, predictive capacity. 

Another interesting conclusion arising from Table~\ref{tab:AUC_Summary} regards the usefulness of using multiple networks to calculate centrality measures, even without using a more sophisticated propagation procedure, such as the multilayer Personalized PageRank. The Baseline + Centrality (single networks) model which includes the extracted centrality variables  of each independent network, is within one percent point of improvement compared to the model including all network variables in both Baseline and Behaviour benchmarks (11.1\% vs 11.7\% in the Baseline model and  3.3\% vs 3.4\% in the Behaviour model). This hints that just the use of multiple networks can bring benefits to modellers, but those seeking to extract every point of accuracy will be better served by the more sophisticated multilayer personalized PageRank approach.

\begin{figure}[tb!]
    \centering
    \includegraphics[width=\textwidth]{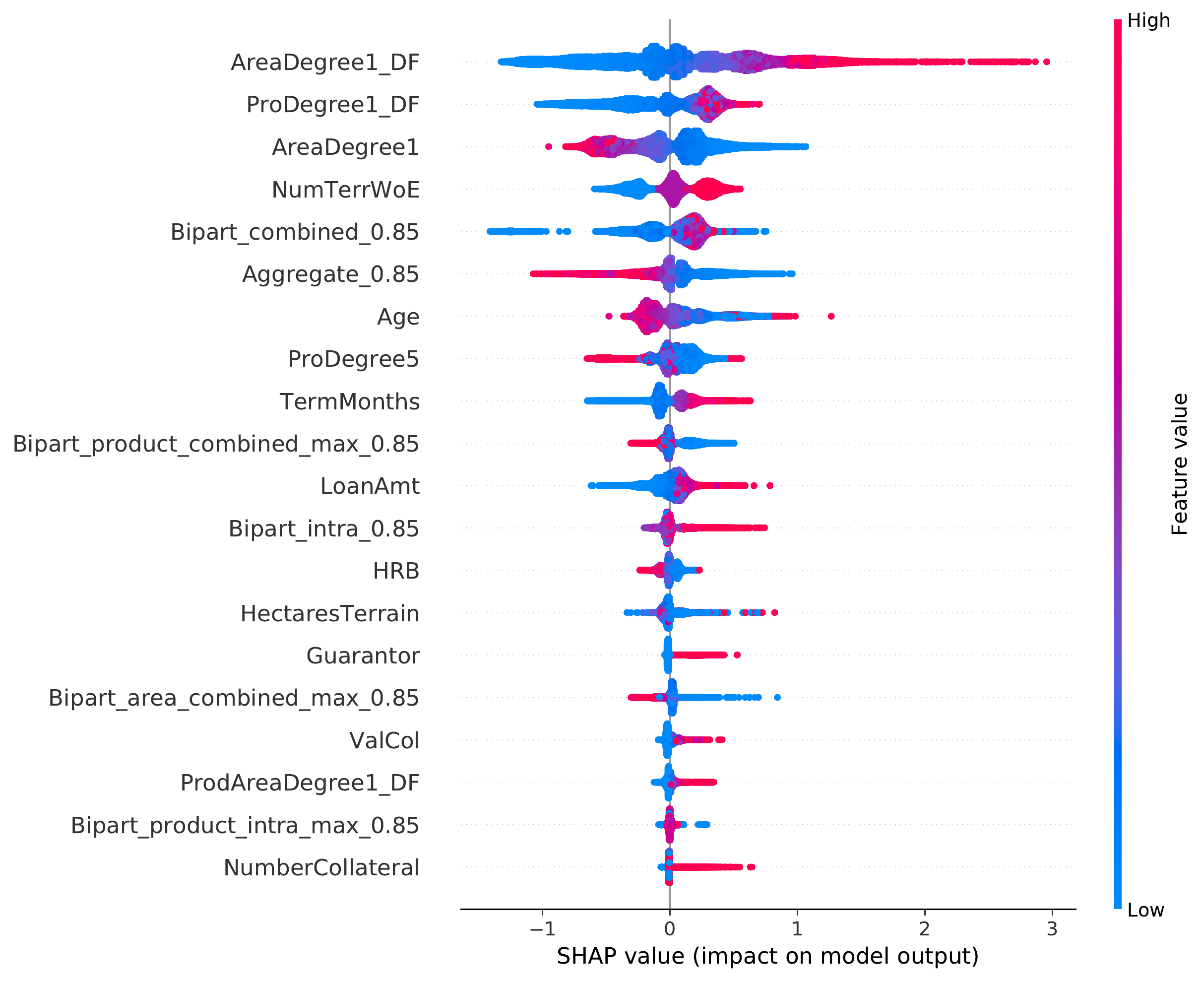}
    \caption{Summary of average Shapley value for a subset of cases. Order by variable importance.}
    \label{fig:ShapleyImportance}
\end{figure}

To dig deeper into the attribute importance, we have used the TreeSHAP method \citep{shap2017} to calculate the average Shapley value of each variable over a subset of the data set, for the Baseline + All Network Variables model. This allows us to create Figure~\ref{fig:ShapleyImportance} which shows the most predictive variables (from top to bottom in order of importance) but also the value the variables take, from lower values in blue to higher values in red. We can also see how these affect prediction, as a highly positive (negative) value in the horizontal axis is associated with a highly positive (negative) effect in the probability of default.

The very first conclusion is that the network variables dominate the most important features. The centrality variables for defaulters (showing how ``close'' borrowers are to previously defaulted cases) are the most significant variables in the model. This is very strong evidence to the existence of correlated default: the more connected a case is to other defaulters in the network translates to a greatly increased risk of default. This effect, for the agricultural lending case we are studying, is stronger in the area network than the product network, also showing events over a geographical area are more significant than events affecting a unique product. In this context: a drought is more significant than a plague. This previously unknown information can be very useful to design preemptive measures to mitigate losses in catastrophic events, and show the importance of a multilayer network approach. Regarding the multilayer PageRank variables, the most complete one (Bipart\_combined) is the most predictive variable. This hints that the multilayer network allows for a richer propagation of risk. This is followed by the Aggregate variable, which is a PageRank calculation over the ``aggregated'' network, which also hints as network connectivity as the source of predictive power.

We have established that network variables are significant and predictive. The next section studies in depth the behaviour of the most significant network variables and the takeaways that can be extracted.

\begin{figure}[t!b]
    \centering
    \begin{subfigure}[b]{0.37\textwidth}
        \includegraphics[width=\textwidth]{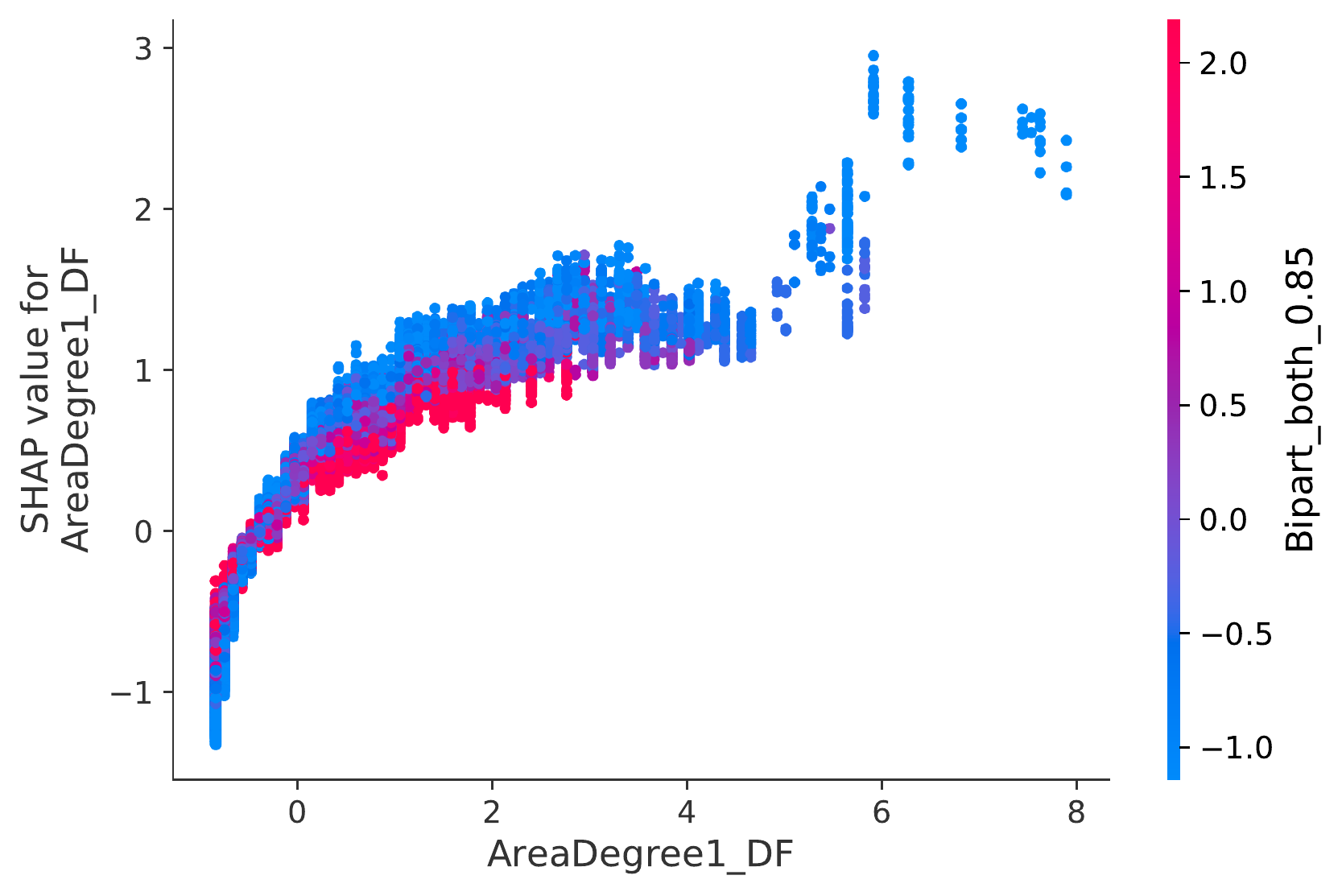}
        \caption{\texttt{AreaDegree1\_DF} (Area centrality with defaulted cases)}
        \label{fig:DepAreaDegree1_DF}
    \end{subfigure}
    \quad 
    \begin{subfigure}[b]{0.37\textwidth}
        \includegraphics[width=\textwidth]{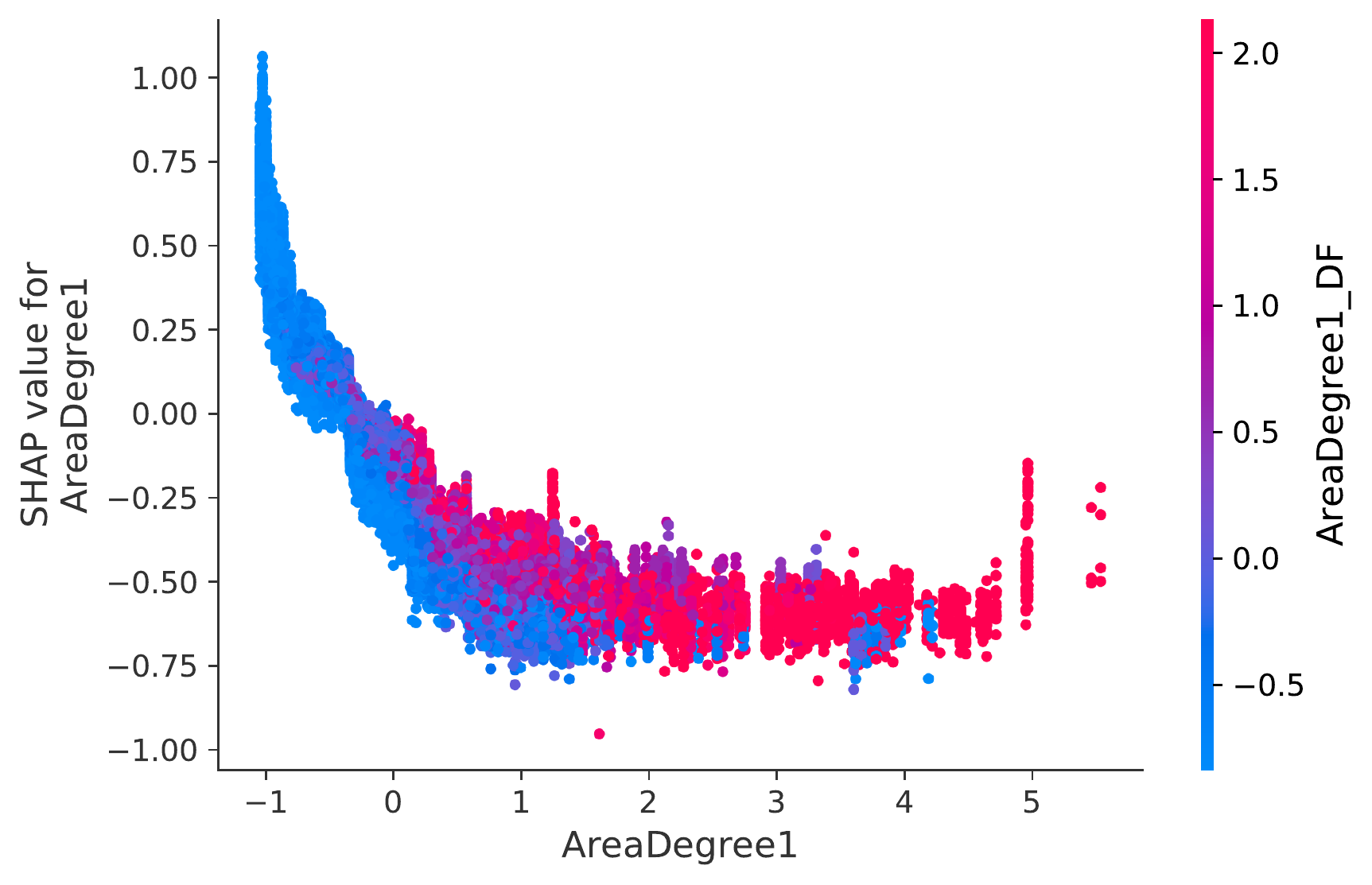}
        \caption{\texttt{AreaDegree1} (Area Centrality, one year)}
        \label{fig:DepAreaDegree1}
    \end{subfigure}
    \newline
    \begin{subfigure}[b]{0.37\textwidth}
        \includegraphics[width=\textwidth]{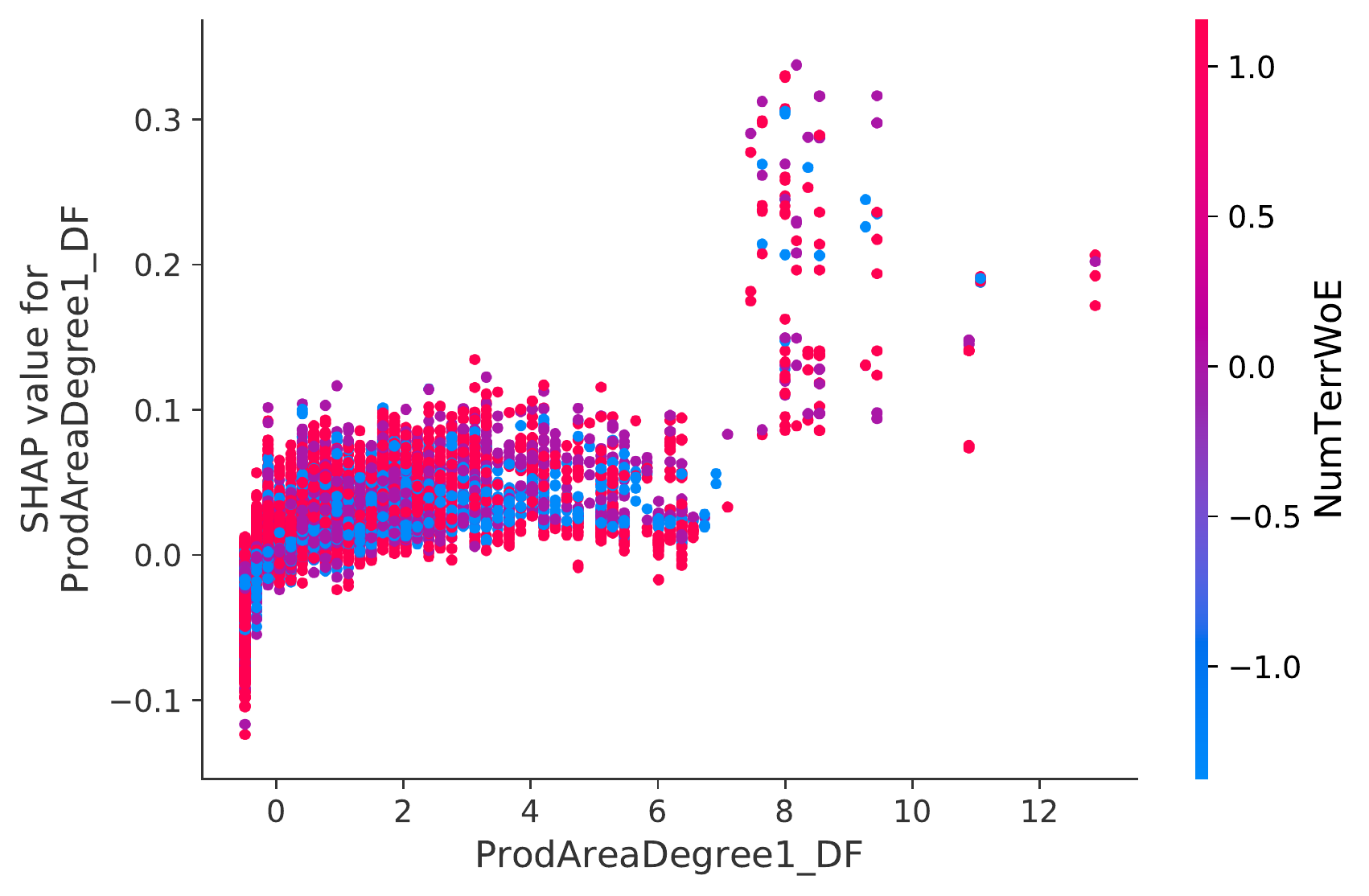}
        \caption{\texttt{ProdAreaDegree1\_DF} (Multilayer centrality with defaulters)}
        \label{fig:DepProdAreaDegree1_DF}
    \end{subfigure}
     \quad 
    \begin{subfigure}[b]{0.37\textwidth}
        \includegraphics[width=\textwidth]{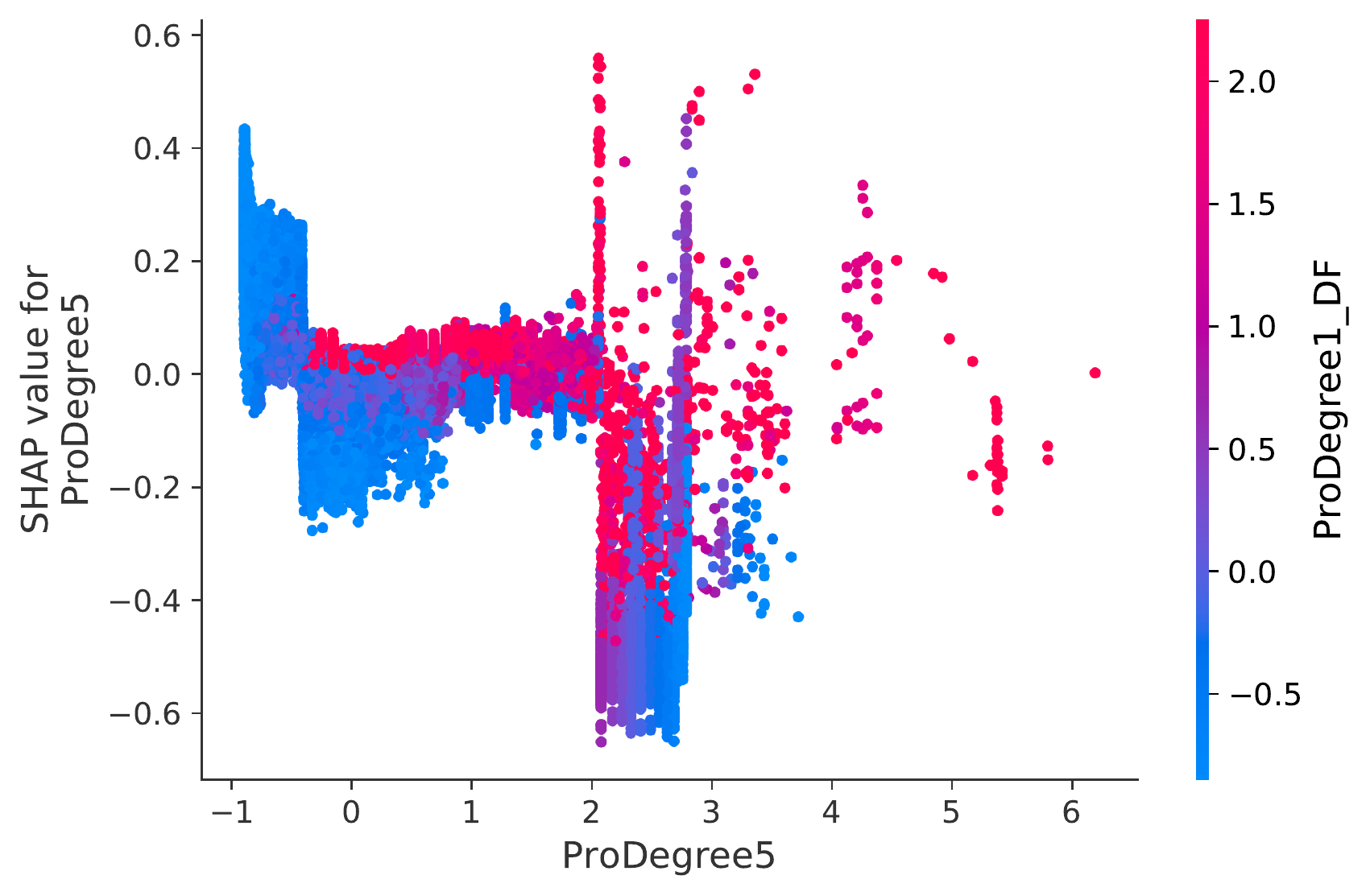}
        \caption{\texttt{ProDegree5} (Product centrality, 5 years)}
        \label{fig:DepProDegree5}
    \end{subfigure}
    \newline
    \begin{subfigure}[b]{0.37\textwidth}
        \includegraphics[width=\textwidth]{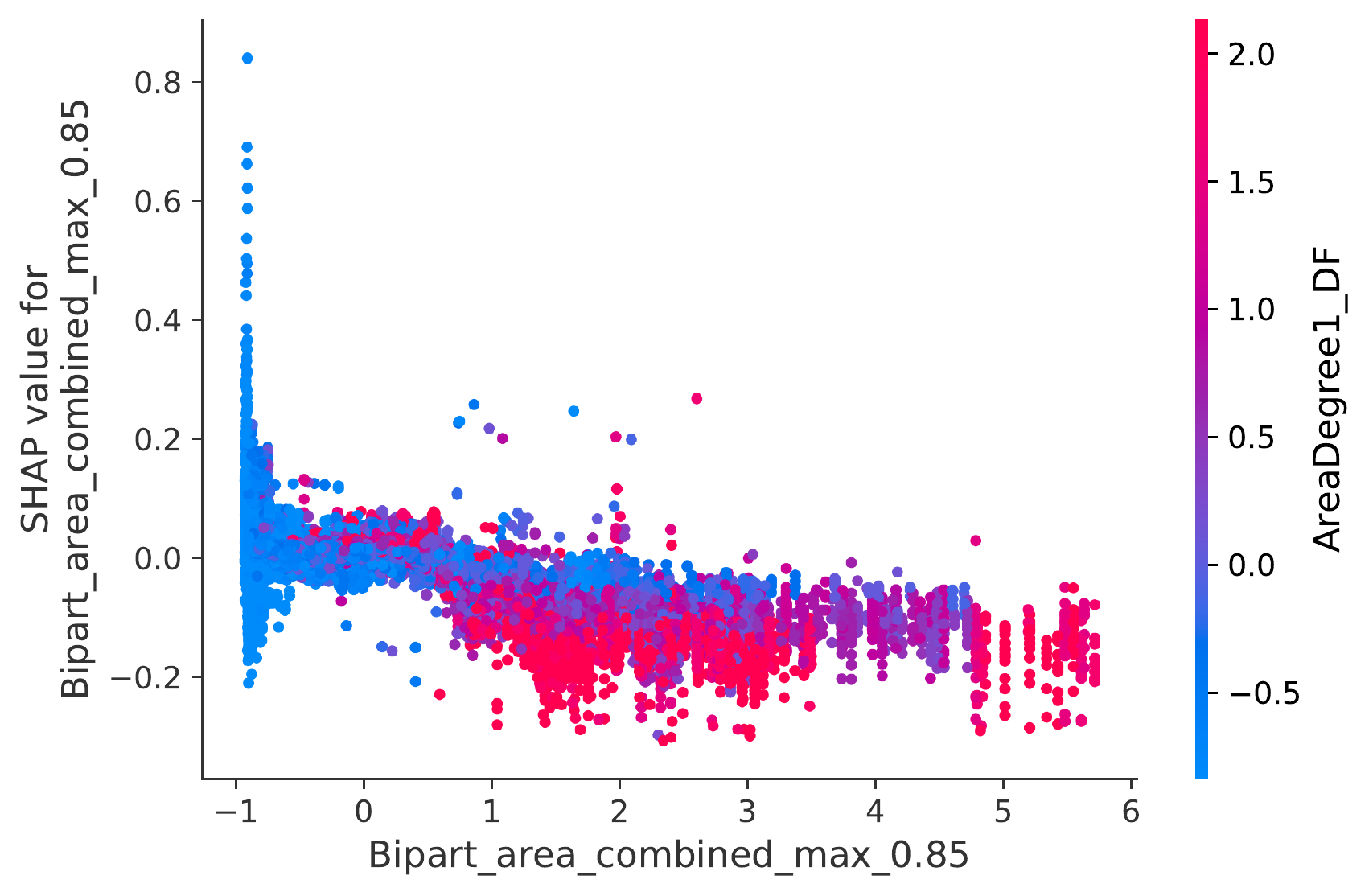}
        \caption{\texttt{Bipart\_area\_combined} (Pagerank score over the area network)}
        \label{fig:DepBipart_area_both}
    \end{subfigure}
    \quad 
    \begin{subfigure}[b]{0.37\textwidth}
        \includegraphics[width=\textwidth]{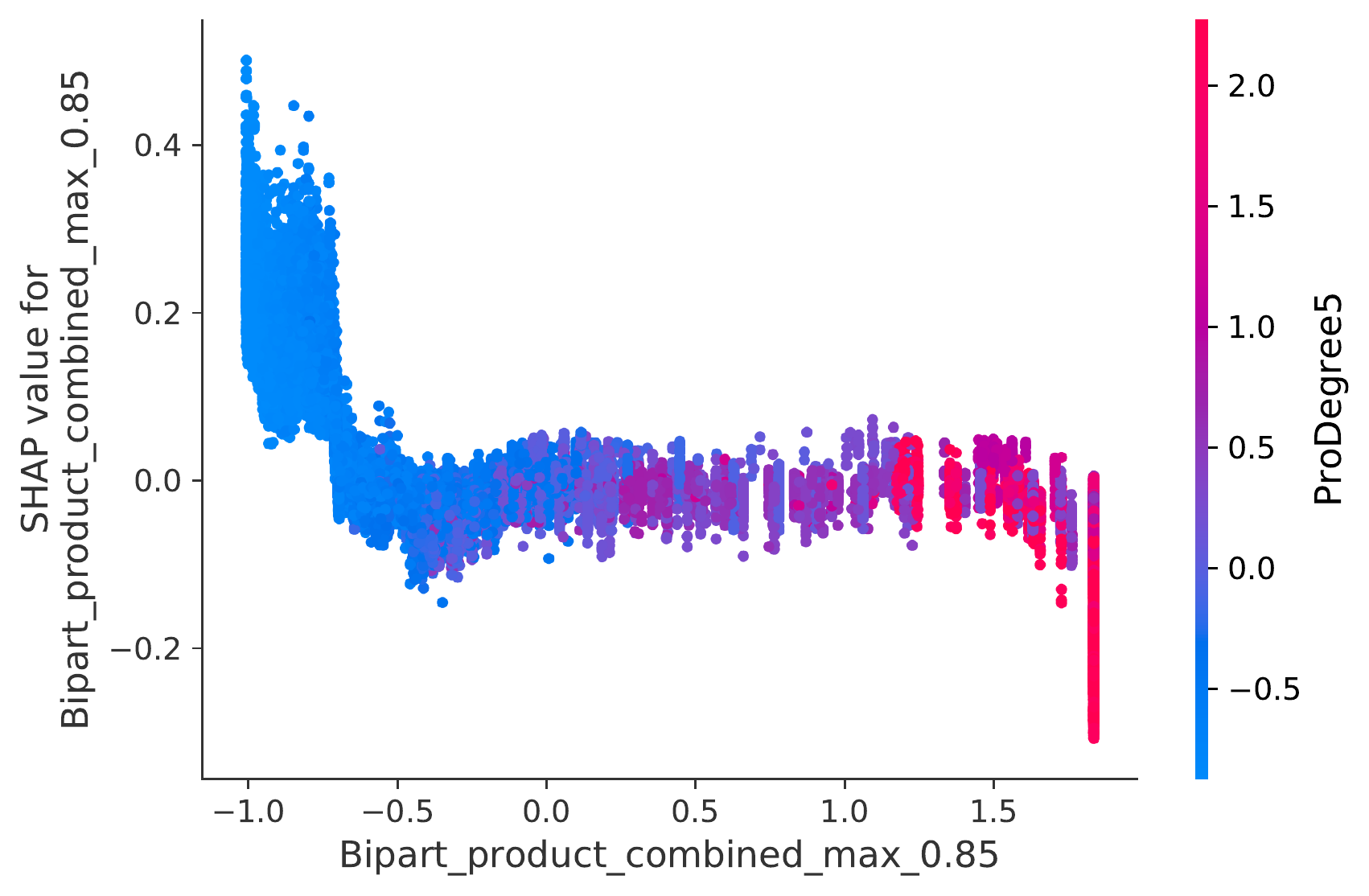}
        \caption{\texttt{Bipart\_product\_combined} (Pagerank score over the product network)}
        \label{fig:DepBipart_product_both}
    \end{subfigure}
    \newline
    \begin{subfigure}[b]{0.37\textwidth}
        \includegraphics[width=\textwidth]{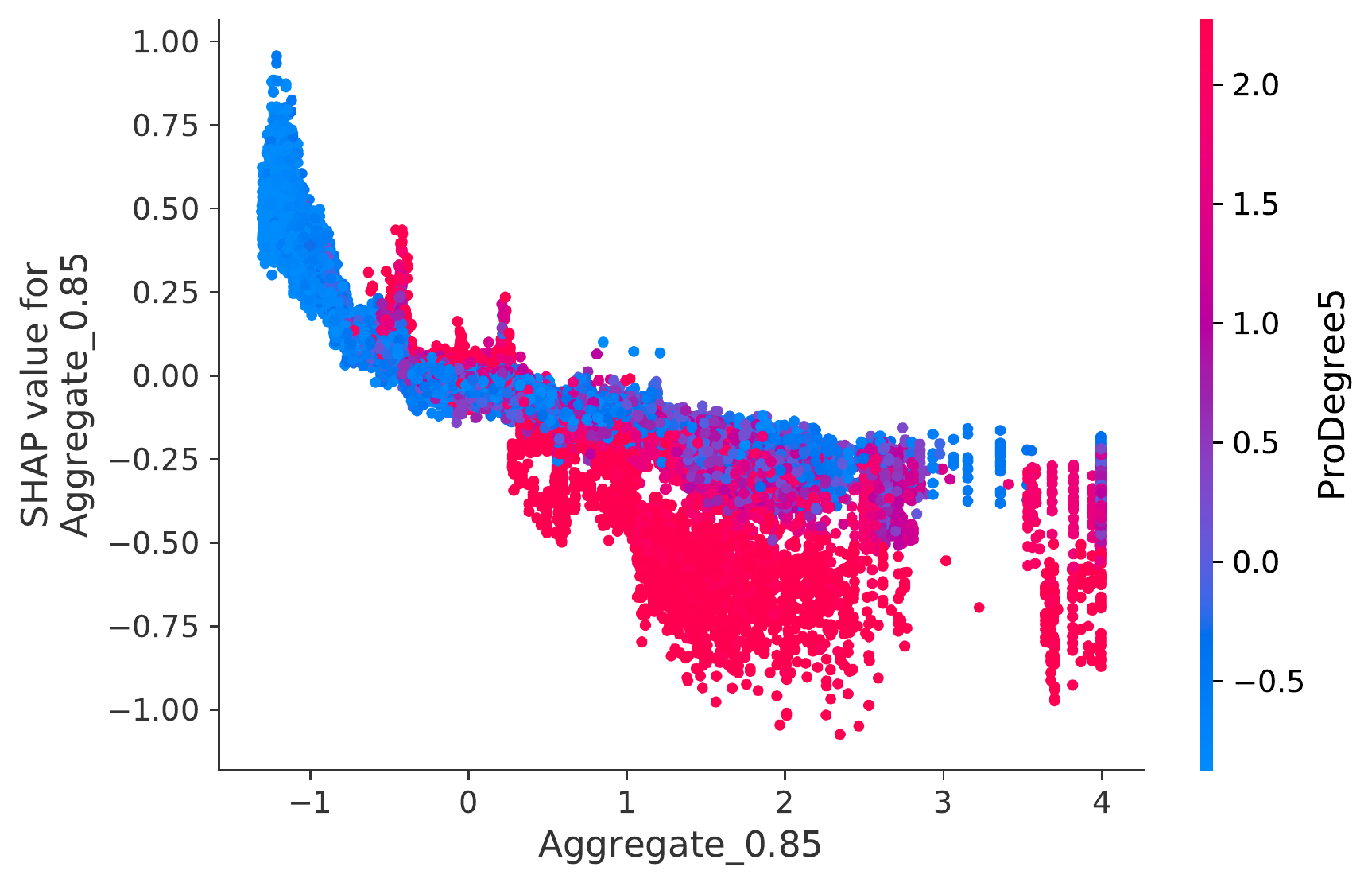}
        \caption{\texttt{Aggregate} (Flattened Pagerank score)}
        \label{fig:DepAggregate}
    \end{subfigure}
    \quad
    \begin{subfigure}[b]{0.37\textwidth}
        \includegraphics[width=\textwidth]{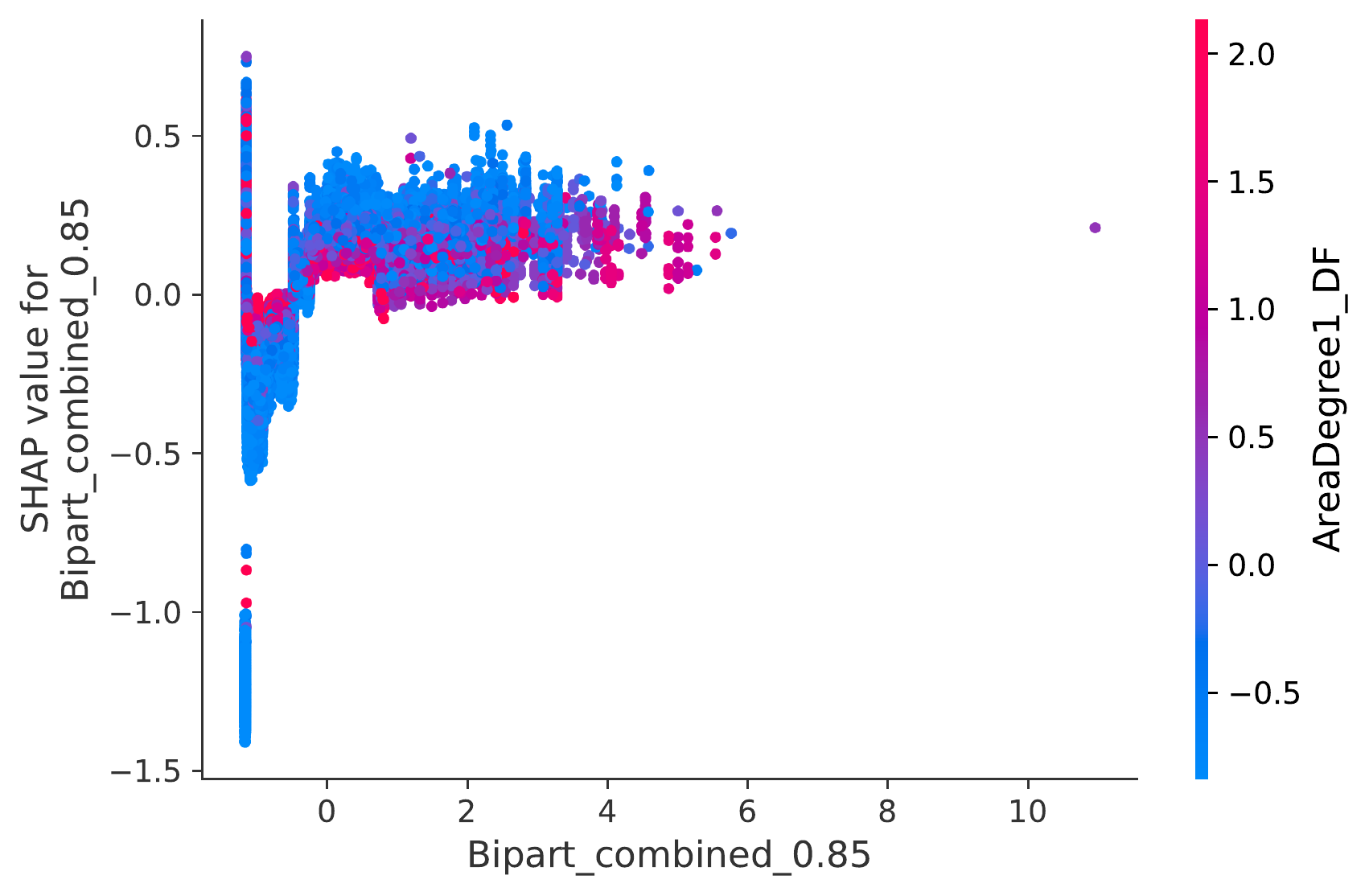}
        \caption{\texttt{Bipart\_combined} (Multilayer pagerank score)}
        \label{fig:DepBipart_both}
    \end{subfigure}
    \caption{Dependency plots for network variables. The colour shows the value of the closest variable by correlation.}\label{fig:Dependency}

\end{figure}

\subsection{Model Interpretation}

For this section we will analyze how the key network variables interact in the model to make predictions. For this, we will use TreeShap's Dependency Plots over the Baseline + All Network Variables XGBoosting model. The dependency plots allow studying the impact of each variable on the prediction (using the Shapley values to study how they influence a prediction towards a value of 0 or 1), and also study how each variable relates to the values of the closest predictor (in terms of correlation). This allows us to paint a very detailed image of how a prediction is calculated.

The first interesting result is that most network effects are markedly non-linear. Extreme values in the variables cause strong movements in the predictions. This means that highly (slightly) connected borrowers are at higher (lower) risk of default, but moderately connected ones in general do not have such a strong network risk.

Focusing on particular segments of variables, the first and second rows (Figures~\ref{fig:DepAreaDegree1_DF}, \ref{fig:DepAreaDegree1}, \ref{fig:DepProdAreaDegree1_DF} and \ref{fig:DepProDegree5}) show a very interesting combined behaviour. The first two present complementing ones, with AreaDegree1\_DF rapidly increasing default risk as the value of the variable increases, which is contrasted by a decreasing risk as the value of the the opposing variable (AreaDegree1) increases. This speaks of a very interesting result: in general, highly connected individuals present \emph{better} behaviour than the average borrower, but individuals highly connected to previously defaulted borrowers are at a higher risk. Figures~\ref{fig:DepProdAreaDegree1_DF} and \ref{fig:DepProDegree5} seem to suggest that the effect of overall centrality is stronger than the centrality with respect to defaulters, thus a stronger community supports individuals more than what a propagated downturn exposes them to risk.  The Shapley values also suggest that a very high connection to defaulters is very predictive of default\footnote{It seems that, at least empirically and for this particular problem, the old adagio ``friends are worth more than money'' seems to hold... unless most people around you are defaulters.}. These two centrality variables also presents a highly non-linear behaviour and for extreme values of the variables they also show strong predictive ability.
 
The third and fourth row of figures show the impact of the PageRank scores. Again, their behaviour is non-linear and values around the centre of the distribution provide little information, with the extremes being better predictors of default or non-default depending on which extreme the variable is located. The PageRank score of the aggregated network seems to be a poor predictor at average values, all scores (and in particular the multilayer score, Figure~\ref{fig:DepBipart_both}) seem to be  excellent predictors of good repayment at low values. As the PageRank scores are calculating exposure to risk, the borrowers who present the lowest probability of default are those that after propagation do not appear to be exposed.

Summarizing both analysis leads to an interesting finding: Centrality shows that how connected a borrower is to defaulters is a good predictor of their default risk, while propagation scores show that even though they might be connected, if their proximity to defaulters is far enough (as measured by the PageRank score) then their risk remains low. It is in using both sets of variables combined where the new predictive capability arises.

\section{Conclusions}
In this paper, we presented a framework for turning a data set into a multilayer network with bipartite intra layer networks, using two or more connector variables in the data set.  This framework could be applied to any data set with reasonable connector variables to obtain an interconnected multilayer network for a data set that is seemingly without network structure. In addition, we developed a personalized multilayer PageRank centrality measure that can be used to propagate influence --some other effect-- from a set of source nodes across the network and its layers, and thus investigate how exposed the nodes are to the influence. 
Finally, we applied our framework and centrality measure to an agricultural lending data set, in which we connected borrowers using variables describing products and districts. We created credit scoring models with variables from the network in addition to traditional variables and saw a significant increase in performance. 

We used our novel personalized multilayer PageRank centrality measure to compute scores that quantify how exposed the nodes are to the influence from the source nodes.  As the scores are computed for all the nodes in the network, the riskiness of both the observations and the connector variables, in this case the products and districts, can be assessed \citep{bravo2020a}.
Furthermore, based on the network and the PageRank scores, we could extract a variety of variables from the network, such as the number of borrowers in the same district and the PageRank score of neighbouring nodes. These variables are properties of the nodes themselves, which we subsequently added to the loan data set and thus enriched it with information from the network. The connector variables in our case were categorical with hundreds of levels, which makes them cumbersome to include in the credit scoring model directly. Through our network construction their effect could be seamlessly included.

An extensive experiment shows that in general network variables are helpful in predicting default, and that multilayer approaches are more useful than single layer ones. We achieve uplifts upward of 10\% on AUC measures over baseline models. Interestingly, the improvements seem to be highly non-linear, suggesting more powerful machine learning approaches must be used to extract optimal predictive capabilities from this information. Finally, and aligned with the literature, behavioural variables are the most powerful predictors of default, but still uplifts of around 2.5\% are attainable by using network variables.

The use of interpretability measures over the network variables also brings interesting takeaways which would not be revealed were these variables not included. The network variables present non-linear, complementary effects which must be carefully studied to understand how default risk propagates across the networks and exposes healthy borrowers to risk. This information can be useful to lenders and regulators to take preemptive measures when they see deteriorating portfolios: at-risk borrowers can be contacted with refinancing offers or support programs can be set up in particularly trying economic conditions.

We believe the results on this work show strong evidence of both correlated default risk, and how this default risk can be explicitly captured by using multilayer bipartite networks. These measures can be used in future works to estimate detailed predictive and econometric models over retail borrowers in many areas.
Our work offers other directions for future research. 
Firstly, there has been a recent advancement of deep learning methods for networked data, called graph neural networks. It would be interesting to develop such a method for multilayer bipartite networks that can learn credit scores --or any node label-- directly from the network structure, and investigate how it compares to the method presented in the current paper.
Secondly, the multilayer PageRank centrality measure can be studied in further detail. This includes the effect of the restart value, the stickiness as well as representation of the source of influence on the node ranking. In this paper we applied it to a network with two layers, but with more layers the complexity increases, which could be investigated.
In addition, the measure could be stress tested in terms of credit risk.
Finally, we would like to apply these methods in other areas beyond credit risk, in particular to use the framework for building multilayer networks with various data sets and to measure Personalized PageRank centrality in them. This would contribute to the field of network science by extending it to settings where network structure is not explicit and allow others to reap the benefits of network science.  

\section{Acknowledgements}

The second author acknowledges the support of the Natural Sciences and Engineering Research Council of Canada (NSERC) [Discovery Grant RGPIN-2020-07114]. This research was undertaken, in part, thanks to funding from the Canada Research Chairs program.


\appendix
\counterwithin{table}{section}
\section{Network variables}\label{app:features}
\begin{table}[h!]
    \scriptsize
    \caption{List of variables extracted from the networks.}
    \label{T:features}
    \centering
    \begin{tabular}{lp{12cm}}\toprule
        Variable & Description  \\ \midrule
 
        \texttt{ProDegree1}&Number of borrowers with the same product in the last one year \\
         \texttt{ProDegree1\_DF}&Number of defaulted borrowers with the same product in the last one year \\
         \texttt{ProDegree5}&	Number of borrowers with the same product in the last five years \\
         \texttt{ProDegree5\_DF}&	Number of defaulted borrowers with the same product in the last five years \\
         \texttt{DistDegree1}	&Number of borrowers in the same district in the last one year \\
         \texttt{DistDegree1\_DF}&	Number of defaulted borrowers in the same district in the last one year \\
         \texttt{DistDegree5}&	Number of borrowers in the same district in the last five years \\
         \texttt{DistDegree5\_DF}	&Number of defaulted borrowers in the same district in the last five years \\
         \texttt{AreaDegree1}&	Number of borrowers in the same area in the last one year \\
         \texttt{AreaDegree1\_DF}&	Number of defaulted borrowers in the same area in the last one year \\
         \texttt{AreaDegree5}&	Number of borrowers in the same area in the last five years \\
         \texttt{AreaDegree5\_DF}&	Number of defaulted borrowers in the same area in the last five years \\
         \texttt{ProdDistDegree1}&Number of borrowers in the same district and with the same product in the last one year\\
         \texttt{ProdDistDegree1\_DF}&	Number of defaulted borrowers in the same district and with the same product in the last one year\\
         \texttt{ProdDistDegree5}&Number of borrowers in the same district and with the same product in the last five years\\
         \texttt{ProdDistDegree5\_DF}&	Number of defaulted borrowers in the same district and with the same product in the last five years\\
         \texttt{ProdAreaDegree1}&Number of borrowers in the same area and with the same product in the last one year	\\
         \texttt{ProdAreaDegree1\_DF}&	Number of defaulted borrowers in the same area and with the same product in the last one year	\\
         \texttt{ProdAreaDegree5}&	Number of borrowers in the same area and with the same product in the last five years	\\
         \texttt{ProdAreaDegree5\_DF}&Number of defaulted borrowers in the same area and with the same product in the last five years	\\
         \texttt{Bipart\_intra}&Personalized PageRank score with the intra influence matrix\\
         \texttt{Bipart\_inter}&Personalized PageRank score with the inter influence matrix\\
        \texttt{Bipart\_combined}&Personalized PageRank score with the combined influence matrix\\
         \texttt{Bipart\_product\_intra\_max}&The personalized PageRank score of the borrower's product with the highest score, computed with the intra influence matrix\\
         \texttt{Bipart\_product\_inter\_max}&The personalized PageRank score of the borrower's product with the highest score, computed with the inter influence matrix\\
         \texttt{Bipart\_product\_combined\_max}&The personalized PageRank score of the borrower's product with the highest score, computed with the combined influence matrix\\
         \texttt{Bipart\_district\_intra\_max}&The personalized PageRank score of the borrower's district with the highest score, computed with the intra influence matrix\\
         \texttt{Bipart\_district\_inter\_max}&The personalized PageRank score of the borrower's district with the highest score, computed with the inter influence matrix\\
         \texttt{Bipart\_district\_combined\_max}&The personalized PageRank score of the borrower's district with the highest score, computed with the combined influence matrix\\
         \texttt{Bipart\_area\_intra\_max}&The personalized PageRank score of the borrower's area with the highest score, computed with the intra influence matrix\\
         \texttt{Bipart\_area\_inter\_max}&The personalized PageRank score of the borrower's area with highest score, computed with the inter influence matrix\\
         \texttt{Bipart\_area\_combined\_max}&The personalized PageRank score of the borrower's area with the highest score, computed with the combined influence matrix\\
        \bottomrule
    \end{tabular}
\end{table}

\newpage
\section{Non-Network Variables Description and Descriptive Statistics}\label{app:descriptive}

\begin{scriptsize}
\begin{table}[h!]
\caption{Variables used in models.}
\label{tab:Descriptive_names}
\begin{tabular}{ll}
\toprule
Variable              & Description                                                               \\
\midrule
Guarantor             & Does the loan have a guarantor?                                           \\
LoanAmt               & Loan amount (masked)                                                      \\
TermMonths            & Term in months                                                            \\
HectaresTerrain       & Surface of terrain where main production occurs                           \\
HRB                   & Return per ha.                                                            \\
Age                   & Age of borrower                                                           \\
NumberCollateral      & Number of securities attached to the operation                            \\
ValCol                & Value of the securities (masked)                                          \\
num\_products         & Number of products the borrower sells                                     \\
Arrears15H            & Percentage of payments in arrears of more than 15 days for previous loans \\
NegativeMoveH         & If any negative movement occurred for previously granted loans            \\
NumOtherLoans         & Number of loans granted before the current one                            \\
NumConcurrLoans       & Number of loans that are still in repayment before the current one        \\
IfOtherLoans          & Does the borrower have any previously granted loan?                       \\
IfConcurrentLoans     & Does the borrower have any concurrently granted loan?                     \\
IfArrear15Current     & Has any of the current loans been in arrears more than 15 days?           \\
MaximumArrearHistoric & Maximum number of days in arrears for all previously granted loans        \\
Default               & Default at 90 days during the life of the loan  \\
\bottomrule
\end{tabular}
\end{table}
\end{scriptsize}

\begin{scriptsize} 
\begin{table}[h!]
\centering
\caption{Descriptive statistics for non-network variables}
\label{tab:Descriptive_stats}
\begin{tabular}{@{}lllll@{}}
\toprule
Variable              & Mean   & Std. Dev. & Min. & Max.  \\ \midrule
Guarantor             & 0.08   & 0.28      & 0    & 1     \\
LoanAmt               & 42.88  & 77.27     & 0    & 7903  \\
TermMonths            & 30.96  & 11.43     & 13   & 71    \\
HectaresTerrain       & 69.54  & 273.67    & 0    & 15411 \\
HRB                   & 4.22   & 5.10      & 0    & 74    \\
Age                   & 52.83  & 13.92     & 17   & 110   \\
NumberCollateral      & 0.50   & 0.53      & 0    & 4     \\
ValCol                & 57.97  & 211.23    & 0    & 27689 \\
num\_products         & 1.52   & 0.84      & 1    & 9     \\
Arrears15H            & 0.19   & 0.29      & 0    & 1     \\
NegativeMoveH         & 0.48   & 0.50      & 0    & 1     \\
NumOtherLoans         & 6.89   & 5.98      & 1    & 55    \\
NumConcurrLoans       & 1.15   & 1.12      & 0    & 11    \\
IfOtherLoans          & 0.89   & 0.32      & 0    & 1     \\
IfConcurrentLoans     & 0.19   & 0.40      & 0    & 1     \\
IfArrear15Current     & 0.15   & 0.36      & 0    & 1     \\
MaximumArrearHistoric & 100.38 & 237.93    & 0    & 3408  \\
Default               & 0.12   & 0.33      & 0    & 1     \\ \bottomrule
\end{tabular}
\end{table}
\end{scriptsize}

\end{document}